\documentclass[vecphys]{svmult}		

\usepackage{makeidx}         
\usepackage{graphicx}        
\usepackage{epsfig}          
\usepackage{multicol}        
\usepackage[bottom]{footmisc}

\makeindex             
                       
\begin{document}
\def\rcgindex#1{\index{#1}}
\def\myidxeffect#1{{\bf\large #1}}
\rcgindex{\myidxeffect{D}!Discrete breathers}
\rcgindex{\myidxeffect{I}!Intrinsic localized modes}
\rcgindex{\myidxeffect{Q}!Quodon gas}
\rcgindex{\myidxeffect{C}!Chemical reaction rates}
\rcgindex{\myidxeffect{I}!Irradiation-induced precipitation}

\title*{THEORY OF A QUODON GAS. WITH APPLICATION TO PRECIPITATION KINETICS IN SOLIDS UNDER IRRADIATION}
\titlerunning{Theory of a quodon gas}
\author{
V. Dubinko
\and
R. Shapovalov
}

\institute{
NSC Kharkov Institute of Physics and Technology, Kharkov 61108, Ukraine \texttt{vdubinko@mail.ru}
}
\maketitle
\abstract
{
Rate theory of the radiation-induced precipitation in solids is modified with account of non-equilibrium fluctuations driven by the ``gas'' of lattice solitons (a.k.a. ``quodons'') produced by irradiation. According to quantitative estimations, a steady-state density of the quodon gas under sufficiently intense irradiation can be as high as the density of phonon gas. The quodon gas may be a powerful driver of the chemical reaction rates under irradiation, the strength of which exponentially increases with irradiation flux and may be comparable with strength of the phonon gas that exponentially increases with temperature. The modified rate theory is applied to modelling of copper precipitation in FeCu binary alloys under electron irradiation. In contrast to the classical rate theory, which disagrees strongly with experimental data on all precipitation parameters, the modified rate theory describes quite well both the evolution of precipitates and the matrix concentration of copper measured by different methods. }


\section{Introduction}
\label{sec:1}

Radiation damage in crystals caused by energetic particles (gamma, electrons,
neutrons, light and heavy ions, etc.) is traditionally characterized by the numbers of point
defects, i.e. vacancies and self-interstitial atoms (a.k.a. Frenkel pairs) and their clusters
produced in displacement events. Their subsequent evolution is governed by diffusion,
which leads to segregation of point defects into vacancy and interstitial clusters, dislocation
loops and voids, a.k.a. \underline{extended defects}. The difference in the ability to absorb point defects
by extended defects is one the main driving force of microstructural evolution under
irradiation. A recovery from radiation damage is traditionally thought to be driven by
thermal fluctuations resulting in the evaporation of single vacancies or atoms from
extended defects. These are examples of Schottky defects, defined as point defects ejected
from an extended defect \cite{Ref1}. Another driver for radiation-induced microstructural evolution
is based the forced atomic relocations resulting from nuclear collisions, a.k.a. \underline{ballistic effects} 
\cite{Ref2} that have been taken into account for the explanation of the dissolution of 
precipitates under cascade damage. Later on it was recognized that the so-called ``thermally activated''
reactions may be strongly modified by irradiation resulting in the \underline{radiation-induced production of Schottky defects} \cite{Ref1, Ref3, Ref4, Ref5, Ref6}, which has essentially the same physical 
nature as the ballistic effects \cite{Ref12}, but, in contrast to the latter, it operates under both cascade 
and non-cascade damage conditions, including sub-threshold electron irradiation that does 
not produce stable Frenkel pairs. The underlying mechanisms for these processes are based 
on the interaction of extended defects with unstable Frenkel pairs, focusing collisions 
(a.k.a. focusons) and with \underline{lattice solitons}.  The latter can be mobile, and these are referred
to bellow as {\itshape \underline{quodons}, which are stable quasi-particles that propagate one-dimensionally and transfer energy along the close packed directions of the lattice}. Quodons may have 
more technological significance than focusons due to much longer propagation ranges 
expected from the nonlinear theory and demonstrated experimentally \cite{Ref6,Ref7}. Russell and 
Eilbeck \cite{Ref7} have presented evidence for the existence of quodons that propagate great
distances in atomic-chain directions in crystals of muscovite, an insulating solid with a 
layered crystal structure. Specifically, when a crystal of muscovite was bombarded with 
alpha-particles ($E<1$ keV) at a given point at room temperature, atoms were ejected from  
remote points on another face of the crystal, lying in atomic chain directions at more than 
$10^7$ unit cells distance from the site of bombardment.

There is also evidence that quodons can occur in non-layered crystals of different 
classes, including insulators, semiconductors and metals. Some examples can be found in 
the radiation damage studies in silicon \cite{Ref8} and germanium \cite{Ref9} and in diffusion experiments 
in polycrystalline austenitic stainless steel \cite{Ref10} and single crystals of copper \cite{Ref11}. This
points out to the necessity of the \underline{modification of the chemical rate theory} with account of  
the quodon-induced energy deposition to the reaction area. Accordingly, the rate theory of 
microstructure evolution in solids has been modified with account of the production of 
Schottky defects at surfaces of extended defects due to their interaction with the radiation--induced 
lattice excitations \cite{Ref1, Ref3, Ref4, Ref5, Ref6}. The modified theory predictions include important
phenomena, which have not been properly understood before, such as the irradiation creep 
\cite{Ref1, Ref3}, radiation-induced annealing of voids \cite{Ref1, Ref6}, saturation of the void growth under high dose irradiation \cite{Ref6}, and the void lattice formation \cite{Ref1, Ref5}.

Until recently the evidence for the existence of lattice solitons provided by 
molecular-dynamics simulations was restricted mainly to one and two-dimensional 
networks of coupled nonlinear oscillators with simple ``toy'' potentials \cite{Ref12, Ref13}, whereas  
reports on their observation in three-dimensional systems using "realistic" molecular--dynamics
potentials were scarce and restricted to alkali halide crystals (see e.g. \cite{Ref14}). The
lattice solitons found in these simulations always drop down from the optical band(s) into 
the phonon gap, and hence become unstable. Consequently, it has been assumed that the 
softening of atomic bonds with increasing vibrational amplitude is a general property of 
crystals, and therefore lattice solitons with frequencies above the top phonon frequency 
cannot occur. However, in their recent paper, Haas et al \cite{Ref15} have provided a new insight  
on this problem by demonstrating that the anharmonicity of metals appears to be very
different from that of insulators. The point is that the essential contribution to the screening 
of the atomic interactions in metals comes from free electrons at the Fermi surface. As a 
consequence, the ion-ion attractive force may acquire a nonmonotonic dependence on the 
atomic distance and may be enhanced resulting in an amplification of even anharmonicities 
for the resulting two-body potentials. This effect can counteract the underlying softening 
associated with the bare potentials with a moderate increase of vibrational amplitudes. As a 
result, in some metals, lattice solitons may exist with frequency {\itshape above the top of the 
phonon spectrum}. Using the known literature values of the pair potentials, Haas et al have found 
that in $Ni$ and $Nb$ this condition is fulfilled. Their molecular-dynamics simulations of the 
nonlinear dynamics of $Ni$ and $Nb$ confirmed that high-frequency lattice solitons may exist 
in these metals, and their corresponding energies may be relatively small, starting from 
\underline{threshold energy of $0.2$ eV}, just above the phonon band. These results allow us to look at the modification of the rate theory based on the quodon dynamics from a different angle as 
compared to the one proposed in refs \cite{Ref1, Ref3, Ref4, Ref5, Ref6}. Initially it has been assumed that the main difference between quodons and focusons is in their path lengths. Accordingly, it was postulated that (I) the quodon production rate was equal to the focuson production rate; and (II) similar to focusons, quodons could eject vacancies from extended defects provided that their energy exceeded the vacancy formation energy. This mechanism is close to the 
classical ballistic mechanism of the precipitate dissolution under cascade damage \cite{Ref2}. 
However, in view of the new results \cite{Ref15}, one can assume that irradiation may produce 
quodons with energies almost as low as that of phonons, which are lower than typical 
focuson energies by orders of magnitude. This assumption has important consequences, as 
discussed in the first part of the paper, where we introduce and develop a new concept of 
quodon ``gas'' proposed recently in ref. \cite{Ref16}. The concept is then used to modify a rate 
theory of the radiation-induced growth (or shrinkage) of precipitates of a new phase 
(extended defects) in solids.  In the second part, the modified rate theory is applied to 
modelling of the nucleation and growth of copper precipitates (nano-sized clusters of 
copper atoms) in $FeCu$ binary alloys under electron irradiation, which has been observed 
experimentally and characterized quantitatively by Mathon et al \cite{Ref17}.


\section{Gas of quodons and its effect on reaction rates in solids}
\label{sec:2}

Based on the low threshold energies for the formation of lattice solitons, as compared to those for focusons, we shall state the following. First, the quodon generation rate may be expected to be much higher than that of focusons due to a higher cross-section for the low energy transfer from a projectile to the lattice. Second, a Schottky defect can not be produced in one collision event between an extended defect and a low energy quodon, which can deliver energy in portions small as compared to the Schottky defect formation barrier. So the ballistic mechanism of the Schottky defect production does not seem to be statistically relevant for a description of the effect caused by quodons. In the following sections we will develop a statistical approach to the modification of reaction rates by the quodon ``gas''.

\subsection{Gas of quodons}
\label{sec:2.1}

It is known that even in the case of displacement damage, only a small part of the energy of impinging particles is spent on generation of stable Frenkel pairs (that require 20-40 eV each) and their clusters (in energetic displacement cascades), while the major part of energy is dissipated into heat, or in other words, it is spent on generation of phonons.

The \underline{first main assumption} in the present paper is that {\itshape quodons are the transient form of the heat generation under irradiation}, which means that they are constantly generated by irradiation, and subsequently lose energy by generating phonons. Let $K_{q}$  be the average rate of quodon generation (per atomic site per second), which should be proportional to the flux of impinging particles, $F_{irr}$ , and the energy deposition density by one particle, $dE_{irr}/dx$, and inversely proportional to the mean quodon energy, 
$\langle E_q \rangle$:  
\begin{equation}\label{eq:1}
K_{q}(F_{irr})=
k^{q}_{eff}F_{irr}\left(\frac{dE_{irr}}{dx}\right)\frac{w}{\langle E_q \rangle},
\end{equation}
where $w$ is the atomic volume, and  $k^{q}_{eff}$ is the quodon production efficiency that depends on material and irradiation conditions and can range from zero (no quodon generation) to unity (e.g. under sub-threshold irradiation that does not produce stable defects).

Then the mean density of quodon gas under steady-state irradiation (the number of quodons per unit volume) will be given simply by the product of their mean generation rate and the life-time, $\tau_q$:
\begin{equation}\label{eq:2}
N_q\left(F_e,T_{irr}\right)=\frac{K_q\left(F_e,E_e\right)\tau_q}{w},\quad
\tau_q(T)=\frac{l_q(T)}{c_q},
\end{equation}
where $c_q$ is the quodon propagation speed, which is assumed bellow to be close to the sound velocity, $c_s$, and $l_q(T)$ is the quodon propagation range before decay.

From a quantum theory of intrinsic localized modes \cite{Ref14} it is known that they decay by generating phonons. Similar to that, quodons are assumed lose energy in quodon-phonon collisions {\itshape by portions (or quanta)} ${\rm\Delta} E_{qp}$. Then the mean quodon propagation range may be written as
\begin{equation}\label{eq:3}
l_q(T,\langle E_q \rangle )=\frac{\langle l_{qp}(T) \rangle}{{\rm\epsilon}_{pq}},\quad
{\rm\epsilon}_{pq}\equiv \frac{{\rm\Delta} E_{qp}}{\langle E_q \rangle},
\end{equation}
where $\langle l_{qp}(T) \rangle$ is the mean length of quodon free path between collisions with phonons, which  is determined by a well known formula for a 1--D propagating particle in a medium with scattering centers of a given density and cross-section:
\begin{equation}\label{eq:4}
\langle l_{qp}(T) \rangle=\frac{1}{\displaystyle \pi R^{2}_{qp}\langle N_p(T) \rangle},\quad
\langle N_p(T) \rangle=
\frac{1}{w}\frac{1}{\displaystyle \exp\!\left(\frac{\hbar \omega_D}{k_B T} \right) -1},
\end{equation}
where $R_{qp}$ is the effective quodon-phonon cross-section radius, $\langle N_p(T) \rangle$ is the density of phonons having a high-frequency $\omega_D$ (a.k.a. Debye phonons), $k_B$ is the Boltzmann constant and $T$ is the absolute temperature. We assume here for simplicity that Debye phonons are the main contributors to the quodon decay due to the loss of energy in each collision between them represented by the energy coupling parameter 
${\rm\Delta}E_{qp} \approx \hbar \omega_D$.

These and other material parameters used in the calculations are listed in Table~\ref{tab:Param}.

\begin{figure}
\center
\includegraphics[width=11.5cm]{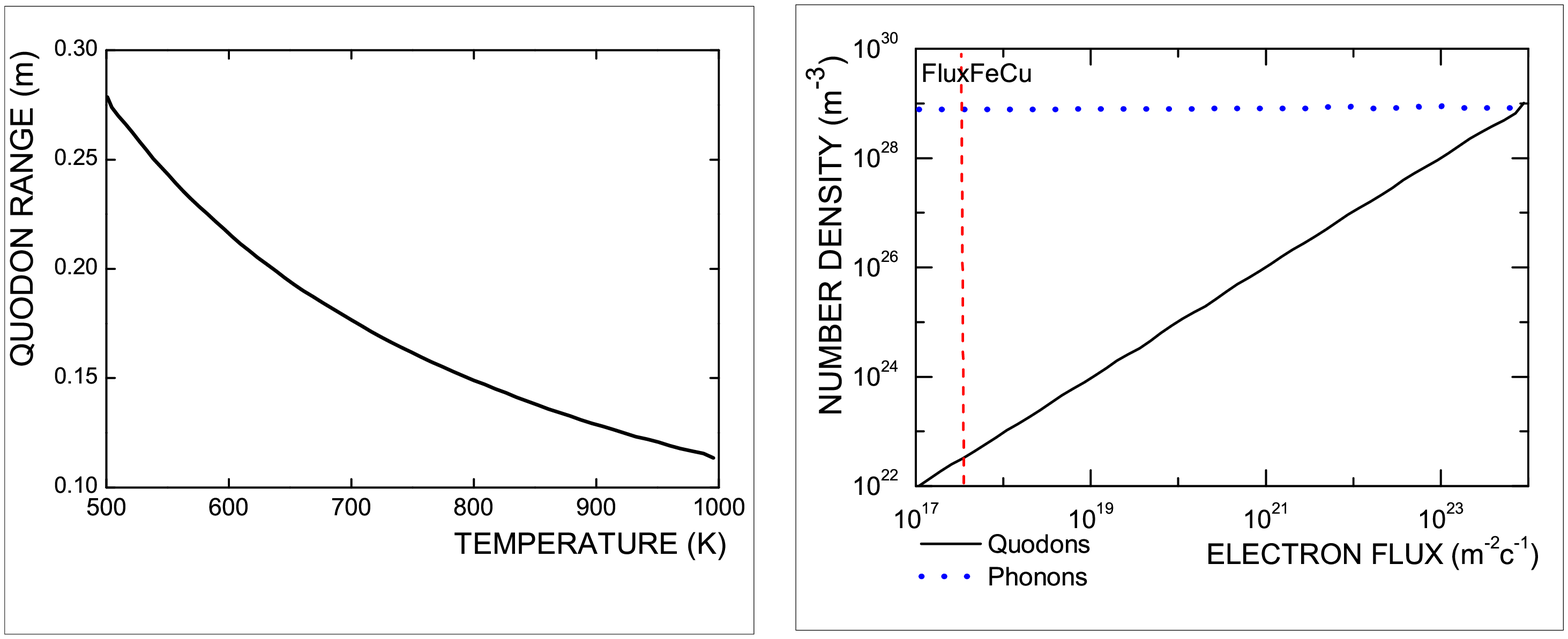}
\caption{The temperature dependence of the quodon propagation range $l_q(T)$ and the density of quodon gas vs. electron flux, $F_{irr}=F_e$ at the irradiation temperature 563 K. The vertical dotted line corresponds to irradiation flux \cite{Ref17}. Density of the gas of Debye phonons at 563 K is shown for comparison.}
\label{fig:1}  
\end{figure}

As can be seen, the quodon-phonon cross-section radius is extremely small ($\sim10^{-12}$ m), and so the quodon propagation range can reach tens of centimeters (Fig.~\ref{fig:1}), which agrees with experimental data on tracks produced by quodons in mica muscovite \cite{Ref7}. This enormous ranges in real crystals that contain structural defects can be understood only assuming that quodons can both lose and gain energy in the scattering process with extended defects. So our second main assumption is that in the collision events between quodons and extended defects, quodons lose and gain energy with equal probability.
\begin{table}
\center
\caption{Material ($Fe$) and irradiation parameters used in calculations.}
\label{tab:Param}      
\begin{tabular}{ll}
\hline\noalign{\smallskip}
Parameter & \quad Value  \\
\noalign{\smallskip}\hline\noalign{\smallskip}
Atomic spacing, $b$, $\rm m$ 																	& \quad $2.96\times 10^{-10}$ \\
Atomic volume of the host lattice, $w$			 									& \quad $0.5\times b^3$       \\
Sound velocity, $c_s$, $\rm m/s$                              & \quad $3.82\times10^3$      \\
Maximum phonon frequency in $Fe$ and $Cu$,
$\omega^{Fe, Cu}_D$, $\rm s^{-1}$															& \quad $6\times10^{13}, 7\times10^{12}$ \\
Quodon-phonon coupling,
${\rm \Delta}E_{qp} \approx \hbar \omega^{Fe}_D$, $\rm eV$    & \quad $0.207$ 							\\
Quodon-phonon cross-section, $R_{qp}$, $\rm m$								& \quad $9.06\times10^{-13}$  \\
Quodon-defect coupling,
$V_q \approx \hbar \omega^{Cu}_D$, $\rm eV$										& \quad $0.029$               \\
Defect energy relaxation time, $\tau_0$, s										& \quad $8\times10^{-8}$			\\
Mean quodon energy, $\langle E_q\rangle$, $\rm eV$						& \quad $1$										\\
Irradiation temperature \cite{Ref17}, $T_{irr}$, $\rm K$			& \quad $563$									\\
Quodon life-time, $\tau_q \left( T_{irr}\right)$, $\rm s$			& \quad $7.184\times10^{-5}$  \\
Electron flux \cite{Ref17}, $F_e$, $\rm m^{-2}s^{-1}$					& \quad $4\times10^{17}$			\\
Electron energy \cite{Ref17}, $E_e$, $\rm MeV$								& \quad $2.5$									\\
Displacement energy \cite{Ref17}, $E_d$, $\rm eV$							& \quad $30$									\\
Displacement rate \cite{Ref17}, 
$K_d\left( F_e, E_e\right)$, $\rm s^{-1}$											& \quad $2.0\times10^{-9}$		\\
Quodon production efficiency, $k^q_{eff}$											& \quad $1$										\\
Quodon production rate \cite{Ref17},
$K_q\left( F_e, E_e\right)$, $\rm s^{-1}$											& \quad $7.27\times10^{-3}$		\\
Quodon propagation length,
$l_q\left( T_{irr}\right)$, $\rm m$														& \quad $0.23$								\\
Migration energy of vacancies, $E_{vm}$, $\rm eV$							& \quad $0.91$								\\
Pre-exponent factor, $D^0_v$, $\rm m^2/s$											& \quad $10^{-5}$							\\
Vacancy formation energy, 
$E^{th}_{vf}$, $\rm eV$																				& \quad $2.0$									\\
$Cu$-vacancy migration energy,
${\rm \Delta}E^m_{Cu, V}$, $\rm eV$														& \quad $0.89$								\\
$Cu$-vacancy binding energy,
$E^b_{Cu, V}$, $\rm eV$																				& \quad $0.6$ 								\\
$Cu$ migration frequency factor,
$\upsilon_{Cu}$, $\rm s^{-1}$																	& \quad $7\times10^{-12}$			\\
$Cu$ migration entropy factor,
${\rm \Delta}S^m_{Cu, V}$																			& \quad $2\cdot k_B$					\\
$Cu$ dissolution energy, $\rm eV$														  & \quad $0.586$								\\
Reaction activation volume, $W_a$															& \quad $10\cdot w$						\\
\noalign{\smallskip}\hline
\end{tabular}
\end{table}

As can be seen from Table~\ref{tab:Param} and Fig.~\ref{fig:1}, the quodon generation rate under irradiation conditions \cite{Ref17} can exceed the displacement rate by 6 orders of magnitude, and the density of quodon gas becomes comparable to that of Debye phonons at the electron flux of $10^{24}m^{-2}s^{-1}$, which corresponds to the displacement rate of $\sim\!5\!\times\!10^{-3}s^{-1}$ that is a typical value for radiation damage studies using electron beams. So we may conclude that under irradiation a crystal contains a mixture of ``gases'' of quasi-particles, namely, almost equilibrium phonons and strongly non-equilibrium quodons, the densities of which may be comparable. After irradiation is switched off, quodons transfer their energy to phonons and disappear over a short relaxation time
$\sim\tau_q\approx7\times10^{-5}~s$, as the crystal attains a thermal equilibrium state. In the following section, we consider the effect of quodon gas on chemical reaction rates under irradiation.
\subsection{Modification of reaction rates in solids under irradiation}
\label{sec:2.2}

The ``gas'' of phonons is responsible for the temperature effect on the reaction rates, $\dot{R}$ , which is expressed by Arrhenius' law:
\begin{equation}\label{eq:5}
\dot{R}=\omega_p \exp\!\left(-\frac{E_a}{k_B T}\right), 
\end{equation}
where $\omega_p$ and $E_a$ are the frequency factor and the activation barrier, respectively. The reaction activation barrier is determined by the maximum {\itshape free energy change} of the system imposed by reaction (a.k.a. the free energy barrier). \underline{Phonons} represent thermal fluctuational forces acting on a system (e.g. point defect and its surrounding) and helping it to overcome the barrier. The strength of the phonon's ``white noise'' is proportional to the temperature and enters the denominator in the exponential argument.

On the other hand, unlike phonons, which are delocalized, quodons are strongly localized on several lattice sites, and so there is no reason to expect the effects on reaction rates of these two very different species to be similar. According to modelling in one-dimensional chains of coupled nonlinear oscillators, in the collision event between two quodons, they behave almost like particles, which undergo elastic and non-elastic collisions, accompanied with a change in momentum and energy. Based on that picture, we assume that in the collision event between a quodon and an extended defect, the latter can gain (or loose) some portion of energy at the place of collision, and so the resulting energy of the system including the extended defect will undergo stochastic deviations from its average value. To include these deviations in the reaction scheme let us recall the thermodynamic perturbation theory developed by Peierls (1932) for systems, the energy of which can not be defined precisely due to small (and hard to detect) deviations from the ground state \cite{Ref18}. Then the total \underline{potential energy} of the system can be written as a sum
\begin{equation}\label{eq:6}
E=E_0 + V,\quad
V << {\rm min}(k_B T, E_0),
\end{equation}
where $V$ is the stochastic deviation of the energy from the ground value $E_0$. The system free Helmholtz energy, $\rm\Phi$, is defined by the integral over the Gibbs ensemble:
\begin{equation}\label{eq:7}
\begin{array}{rcl}
\displaystyle\exp\!\left(-\frac{\rm\Phi}{k_B T}\right) & = & \displaystyle\int\!
{\displaystyle\exp\!\left(-\frac{E_0 + V}{k_B T}\right)d\rm\Gamma} \\[2.0ex]
& \approx &
\displaystyle\int\!
{\displaystyle\exp\!\left(-\frac{E_0}{k_B T}\right)\left(1 - \frac{V}{k_B T}
 + \frac{V^2}{2(k_B T)^2} \right)d\rm\Gamma},
\end{array}
\end{equation}
where $d\rm\Gamma$ is the configuration volume in the phase space. Taking the logarithm of eq.~(\ref{eq:7}) and expanding into the Taylor's series again one gets the free energy in the following form:
\begin{equation}\label{eq:8}
\begin{array}{rcl}
\rm\Phi = \rm\Phi_0 & + & \displaystyle\int\!\left( V - \frac{V^2}{2 k_B T} \right)
\displaystyle\exp\!\left( - \frac{{\rm\Phi_0} - E_0}{k_B T} \right)d\rm\Gamma \\[2.0ex]
& + & \displaystyle\frac{1}{2 k_B T}\left[\displaystyle\int\!V\exp\!\left( - \frac{{\rm\Phi_0} - E_0}{k_B T} \right) d\rm\Gamma \right]^2,
\end{array}
\end{equation}
where $\rm\Phi_0$ is the ground free energy at $V = 0$. The integrals in eq.~(\ref{eq:8}) are the average values of the corresponding functions evaluated using the unperturbed Gibbs distribution function, and so eq.~(\ref{eq:8}) can be written as follows:
\begin{equation}\label{eq:9}
{\rm\Phi} \approx {\rm\Phi_0} + \langle V \rangle - 
\frac{1}{2 k_B T}\left[ \langle V^2 \rangle - {\langle V \rangle}^2 \right],
\end{equation}
where the brackets $\langle\phantom{f}\rangle$ designate the integration over the Gibbs ensemble so that 
$\langle V \rangle$ is simply the average value of the perturbation energy (which vanishes in the case of random fluctuations of alternative signs) while the second term in eq.~(\ref{eq:9}) is \underline{always negative} and it is proportional to the square of the energy deviation dispersion due to random fluctuations of the system energy. 
\begin{figure}
\center
\includegraphics[width=8cm]{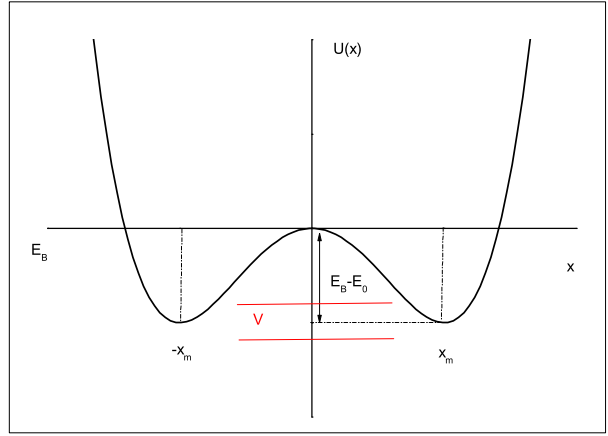}
\caption{Sketch of the double-well \underline{potential landscape} with minima located at $\pm{x_m}$. These are stable states before and after reaction, separated by a potential ``barrier'' with the height changing randomly within the V band induced by quodons.}
\label{fig:2}  
\end{figure}

Now this result can be applied to a system including the ``reaction area'' surrounding a point defect or some region near the surface of extended defect subjected to random collisions with quodons (Fig~\ref{fig:2}). The \underline{free energy} of such system will be decreased according to eq.~(\ref{eq:9}), and so decreased will be the maximum change of the system free energy required for the reaction to occur, i.e. the reaction activation barrier. This decrease has a purely \underline{statistical nature} reflecting the increase of vibrational entropy of the system due to its collisions with quodons.  Accordingly, the Arrhenius' law eq.~(\ref{eq:5}) can be rewritten with account of the energy exchange between quodons and the ``reaction area'':
\begin{equation}\label{eq:10}
\dot{R} = {\rm\omega_0}\exp\!\left(- \frac{E^{q}_{a}}{k_B T} \right),\quad
E^{q}_{a} = \langle E_a \rangle - {\rm\Delta}E_a,\quad
E_a = \frac{\langle V^{2}_{q} \rangle - {\langle V_{q} \rangle}^{2}}{2 k_B T},
\end{equation}
where we have assumed that the average energy of the system will not change due to its collisions with quodons, which can give or take energy with equal probability, i.e. $\langle V_{q} \rangle = 0$.

This result means that under irradiation \underline{all reaction barriers} in a crystal may be decreased by a value that depends on the statistics of the ``gas'' of quodons and their coupling with structural defects. It can be shown that at steady state the energy deviation dispersion in eq.~(\ref{eq:10}) can be expressed via the quodon generation rate $K_q$, the propagation range $l_q$ and the mean portion of energy that is exchanged between a quodon and the reaction area in one collision event, $V_q$ (for Cu precipitates $V_q \approx \hbar \omega^{Cu}_{D}$ - the Debye ``cutoff'' energy of phonons in Cu). To do so let us assume that the system can be found in two states: excited state with energy $\langle E_a \rangle \pm V_q$ due to collisions with quodons and in the ground state with 
energy $\langle E_a \rangle$. The relative fraction of time, in which the system is excited ${{\rm\Delta} t}/t$ is given by the product of the frequency of collisions with quodons, $\omega_q$, and the time it takes to relax to the ground state, $\tau_0$. Then the square of the system energy averaged over large time $t$ (which is equivalent to the averaging over the Gibbs ensemble) can be written as follows:
\begin{equation}\label{eq:11}
\begin{array}{rcl}
\langle V^2_q \rangle & = &
\langle E_a \rangle^2 \left(t - 2 {\rm\Delta}t \right) + \left( \langle E_a \rangle + V_q \right)^2 {\rm\Delta}t -
\left( \langle E_a \rangle - V_q \right)^2 {\rm\Delta}t \\[2.0ex]
& = & \langle E_a \rangle^2 \left(t - 2 \omega_q \tau_0\right)
+ \left( \langle E_a \rangle + E_q \right)^2 \omega_q \tau_0 +
\left( \langle E_a \rangle - V_q \right)^2 \omega_q \tau_0,
\end{array}
\end{equation}
whence it follows that the energy deviation dispersion in eq.~(\ref{eq:10}) is given by
\begin{equation}\label{eq:12}
\langle V^2_q\rangle - \langle V_q \rangle^2 = 2 V^2_q \omega_q \tau_0,
\end{equation}
The frequency of collisions between the reaction area of the radius $R$ and the quodon gas of density $N_q$ can be expressed via the quodon generation rate and propagation range:
\begin{equation}\label{eq:13}
\omega_q = N_q c_q 4 \pi R^2 = \frac{K_q}{w} 4 \pi R^2 l_q,
\end{equation}
Since the quodon generation rate is proportional to the flux of impinging particles $F_{irr}$ (see eq.~(\ref{eq:1})), one can express the modified reaction rate as follows:
\begin{equation}\label{eq:14}
\dot{R} = \dot{R}_0 A\left( f_{irr} \right),\quad
A\left( f_{irr} \right) = \exp\!\left( \frac{k_{mat} F_{irr}}{\left( k_B T \right)^2} \right),\quad
k_{mat}\propto l_q \tau_0,
\end{equation}
where $\dot{R}_0$ is the reaction rate in the ground state (no irradiation) and $A\left( f_{irr} \right)$ is the reaction amplification factor due to the interaction with quodon gas, and $k_{mat}$ is the coefficient determined by quodon statistics and material parameters.

In the following section, the modified reaction rates will be applied to modelling of the second-phase precipitation under irradiation. This phenomenon is of both fundamental and technological importance, since it represents a phase transition of the first order in a strongly non-equilibrium system, i.e. a crystal under irradiation, which changes its service properties as a nuclear material.
%
%
\section{Modeling of the precipitation kinetics under irradiation in the modified rate theory}
\label{sec:3}

Service properties of pressure vessel steels (used in nuclear industry) are very sensitive to precipitation of copper (i.e. aggregation of copper impurities into nano-sized clusters of copper atoms - {\itshape precipitates}), since the precipitates acts as traps for gliding dislocations, which makes the material less ductile and more brittle. These phenomena are called the irradiation hardening and embrittlement of steels \cite{Ref17}. So the kinetics of precipitate nucleation and growth has been the focus of extensive experimental and theoretical studies \cite{Ref19, Ref20, Ref21, Ref22, Ref23}. However, the mechanisms of precipitation under irradiation are still not correctly understood. This is the reason we undertook a modelling of copper precipitation under electron irradiation in dilute (i.e. containing a low concentration of copper) $FeCu$ binary alloys, where experimental trends have been rather comprehensively described by Mathon et al \cite{Ref17}. One of the main advantages of the work \cite{Ref17} is that it describes not only evolution of copper precipitates but also the time dependence of the concentration of copper atoms in the matrix, $\overline{C}_{Cu}(t)$, both under thermal annealing at 773 K and under electron irradiation at 563 K. It appears that $\overline{C}_{Cu}$ at the end of the precipitation process, which corresponds to the copper solubility limit, is of the same order of magnitude in both cases. This result is in a marked disagreement with Arrhenius law that predicts the copper solubility limit at 563 K to be several orders of magnitude lower than observed.

In our view, this contradiction is a very principal one, since the solubility limit value practically determines the rate of the nucleation and growth, and a failure to evaluate it in the classical rate theory makes it impossible to describe correctly the precipitation kinetics under irradiation. To our knowledge, in all up to date models, the solubility limit under irradiation is assumed to be determined only by temperature, similar to the thermal case \cite{Ref20, Ref21, Ref22, Ref23, Ref24, Ref25, Ref26, Ref27}.  As will be demonstrated in the present work, this is a \underline{misleading assumption}.

We present a detailed description of the rate theory scheme for the precipitation kinetics and describe the copper precipitation under thermal annealing and under irradiation in the framework of classical rate theory (Section~\ref{sec:3.1}). In Sections~\ref{sec:3.3}--\ref{sec:3.5}, we modify the rate theory with account of non-equilibrium thermodynamics of ``quodon gas'' and apply it to modelling of copper precipitation in $FeCu$ binary alloys under electron irradiation.

\subsection{Classical rate theory of the precipitation kinetics}
\label{sec:3.1}
As is usual in the classical nucleation theory, $Cu$--precipitates are characterized by a distribution function of its sizes, $f_n$, that is the atomic fraction of precipitates consisting of n-atoms of copper. The time-evolution of the size-distribution function is described by the set of $N-1$ equations (the Becker-Doring {\itshape finite-difference} equation \cite{Ref19}), where $N$ is the maximal size of the precipitates: 
\begin{equation}\label{eq:15}
\frac{\partial f_n}{\partial t} = W^{+}_{n-1}f_{n-1} - W^{-}_{n}f_{n} - W^{+}_{n}f_{n} + W^{-}_{n+1}f_{n+1}, \quad
n = 2,\ldots,N,
\end{equation}
where $W^{+}_{n}$ and $W^{-}_{n}$ are the \underline{forward and backward transition rates}, or the probabilities to increase or decrease a number of atoms in a $n$-atomic precipitate by one per unite time (the reactions involving transitions $f_n \rightarrow f_{n\pm m}$ , where  $m > 1$ are neglected). So the precipitation is considered as a random walk in the size space ($n$) with a step of length $= 1$.

This equation can be transformed into a differential equation. Applying the Taylor's expansion in power series up to the second order one obtains the well known Fokker-Plank {\itshape differential} equation:
\begin{equation}\label{eq:16}
\frac{\partial f_n}{\partial t} = \frac{\partial}{\partial n}\left[\left(W^{-}_{n} - W^{+}_{n}\right)f_{n}\right]
+\frac{\partial^2}{\partial n^2}\left[\frac{1}{2}\left(W^{-}_{n} + W^{+}_{n}\right)f_{n}\right],
\end{equation}

In the adiabatic case (i.e. where the total number of solute atoms is constant), these equations can be integrated numerically provided that the forward and backward transition rates are known. In theory, precipitates are usually assumed to have a spherical form. In this case one has (see e.g. \cite{Ref21})
\begin{equation}\label{eq:17}
W^{+}_{n} = \frac{4 \pi D}{w}R\overline{C}, \quad
W^{-}_{n+1} = \frac{4 \pi D}{w}R C_R,
\end{equation}
where $D$ is the monomer diffusivity, $R$ is the precipitate radius, $overline{C}$ and $C_R$ are the mean monomer concentrations in the bulk (the bar over $C$ designates the average over a macroscopic volume) and the local concentration at the precipitate surface, respectively. The former can be found from the solute conservation low:
\begin{equation}\label{eq:18}
\overline{C} + \sum^{n=N}_{n=2} n f_n = \overline{Q},
\end{equation}
where $\overline{Q}$ is the total volume fraction of solute atoms, and the latter ($C_R$) is usually derived from thermodynamic considerations \cite{Ref17, Ref19}:
\begin{equation}\label{eq:19}
C_R = C^{eq}_1\exp\left(\frac{2\gamma w}{k_B T R}\right), \quad
\gamma = 0.54 \frac{k_B T}{b^2}\left(\ln\frac{1}{C^{eq}_1} - 2 \right),
\end{equation}
where $C^{eq}_1$ is the equilibrium concentration of monomers at a free flat surface, $\gamma$ is the precipitate surface energy according to Cahn-Hilliard theory \cite{Ref17}.

Diffusion of copper in iron occurs by the vacancy mechanism, and so the copper diffusivity $D_Cu$ is proportional to the mean concentration of vacancies in the bulk, $\overline{C}_V$:
\begin{equation}\label{eq:20}
D_{Cu} = \frac{\overline{C}_V}{C^{eq}_V} D^0_{Cu}\exp\left(-\frac{E^m_{Cu}}{k_B T} \right),
\end{equation}
where $E^m_{Cu}$ is the effective migration  energy of copper, $D^0_{Cu}$ is the pre-exponent factor, and $C^{eq}_V$ is the equilibrium concentration of vacancies, which has the classical form 
(\underline{in the case of thermal equilibrium only}), as well as the equilibrium concentration if copper atoms:
%
%
%
%
\begin{eqnarray}
&C^{eq}_V& = C^0_V\exp\left(-\frac{E^f_V}{k_B T}\right), \quad \phantom{f}
C^0_V = \exp\left(\frac{{\rm \Delta}S_V}{k_B} \right),\label{eq:21}\\ 
&C^{eq}_{Cu}& = C^0_{Cu}\exp\left(-\frac{E^f_{Cu}}{k_B T}\right), \quad
C^0_{Cu} = \exp\left(\frac{{\rm \Delta}S_{Cu}}{k_B} \right),\label{eq:22}  
\end{eqnarray}
where $E^f_V$, $E^f_{Cu}$ are the vacancy and copper formation energies, ${\rm \Delta}S_{V, Cu}$ are the entropy factors. In this way, the rate theory is completed, and we present some results obtained for the copper precipitation under conditions of thermal annealing and under electron irradiation in dilute $FeCu$ binary alloys
($\sim\!1 \, {\rm at}~\%\,Cu$) in comparison with experimental data by Mathon et al \cite{Ref17}. Under thermal annealing one has simply $\overline{C}_V = C^{eq}_V$, and so the precipitation rate is determined only by the copper diffusivity. We have tested several different diffusivities in order to achieve the best agreement with experimental data \cite{Ref17}. They are listed in Table~\ref{tab:Diffus}.

\begin{figure}
\center
\includegraphics[width=8cm]{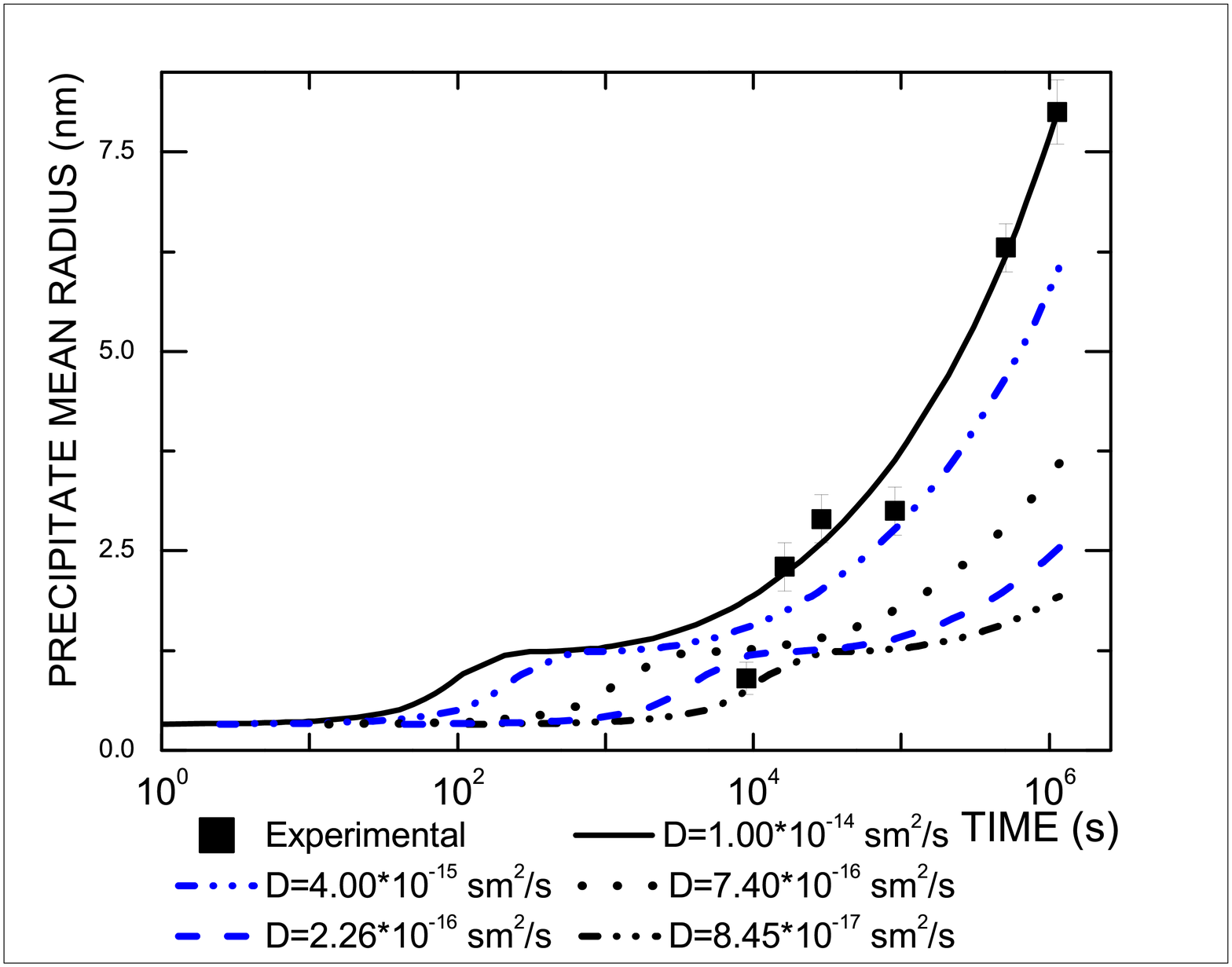}
\caption{Time evolution of the $Cu$ precipitate mean radius under annealing at $773 K$ vs. experimental data \cite{Ref17}.}
\label{fig:3}  
\end{figure}

\begin{figure}
\center
\includegraphics[width=8cm]{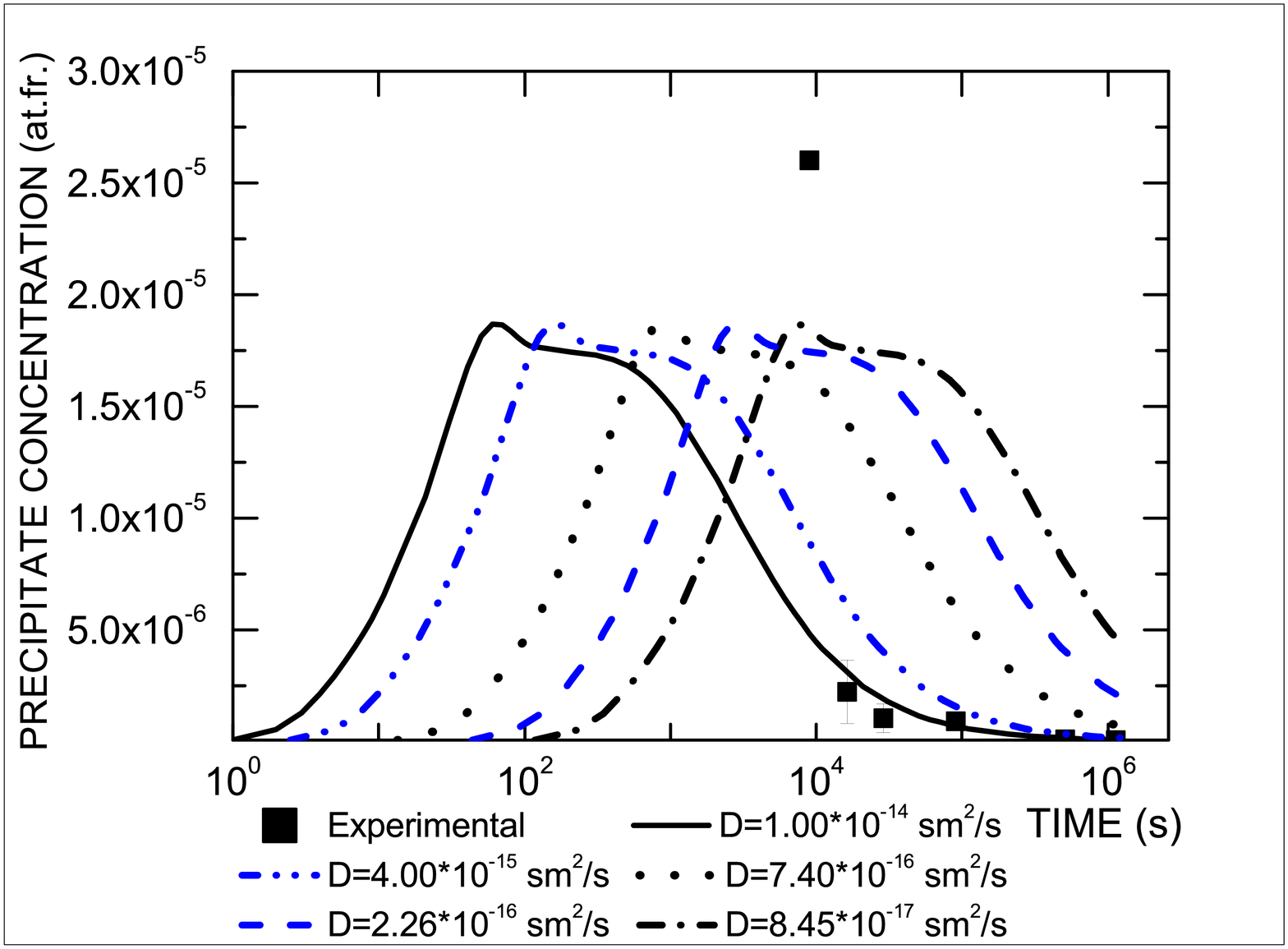}
\caption{Time evolution of the $Cu$ precipitate concentration under annealing at $773 K$ vs. experimental data \cite{Ref17}.}
\label{fig:4}  
\end{figure}

\begin{table}
\center
\caption{Diffusivity of copper in iron given by references listed in the last column.}
\label{tab:Diffus}
\begin{tabular}{lllll}
\hline\noalign{\smallskip}
\# & \quad $D^0_{Cu}, {\rm cm^2/s}$ & \quad $E^m_{Cu}, {\rm eV}$ & \quad $D_{Cu}(773 K), {\rm cm^2/s}$ & \quad Ref \\
\noalign{\smallskip}\hline\noalign{\smallskip}
$1$ & \quad $8.5$ 						 & \quad $2.29$ & \quad $1.00\times10^{-14}$ & \quad This work   \\
$2$ & \quad $3.4$ 						 & \quad $2.29$ & \quad $4.00\times10^{-15}$ & \quad \cite{Ref17}\\
$3$ & \quad $7.2\times10^{-2}$ & \quad $2.29$ & \quad $8.45\times10^{-17}$ & \quad \cite{Ref22}\\
$4$ & \quad $6.3\times10^{-1}$ & \quad $2.29$ & \quad $7.40\times10^{-16}$ & \quad \cite{Ref23}\\
$5$ & \quad $7.08$						 & \quad $2.53$ & \quad $2.56\times10^{-16}$ & \quad \cite{Ref20}\\
\noalign{\smallskip}\hline
\end{tabular}
\end{table}
The results for the mean precipitate radius and the concentration under thermal annealing at 773 K are presented in Figs.~\ref{fig:3}~and~\ref{fig:4}.

It can be seen from these figures that the best fit is achieved using the $Cu$ diffusivity presented in the first line in the Table~\ref{tab:Diffus}, which shows a very good agreement with experimental data in the most reliable region -- at the late stage of precipitation, $t > 1000~{\rm s}$.

Under irradiation, the classical rate theory assumes that the solubility limit of copper \underline{is given by eq.~(\ref{eq:22}) as well as under thermal annealing}, and the only effect of irradiation is the enhancement of copper diffusivity via radiation-induced increase of the mean vacancy concentration in the matrix,
$\overline{C}_V >> C^{eq}_V$. 
\begin{figure}
\center
\includegraphics[width=8cm]{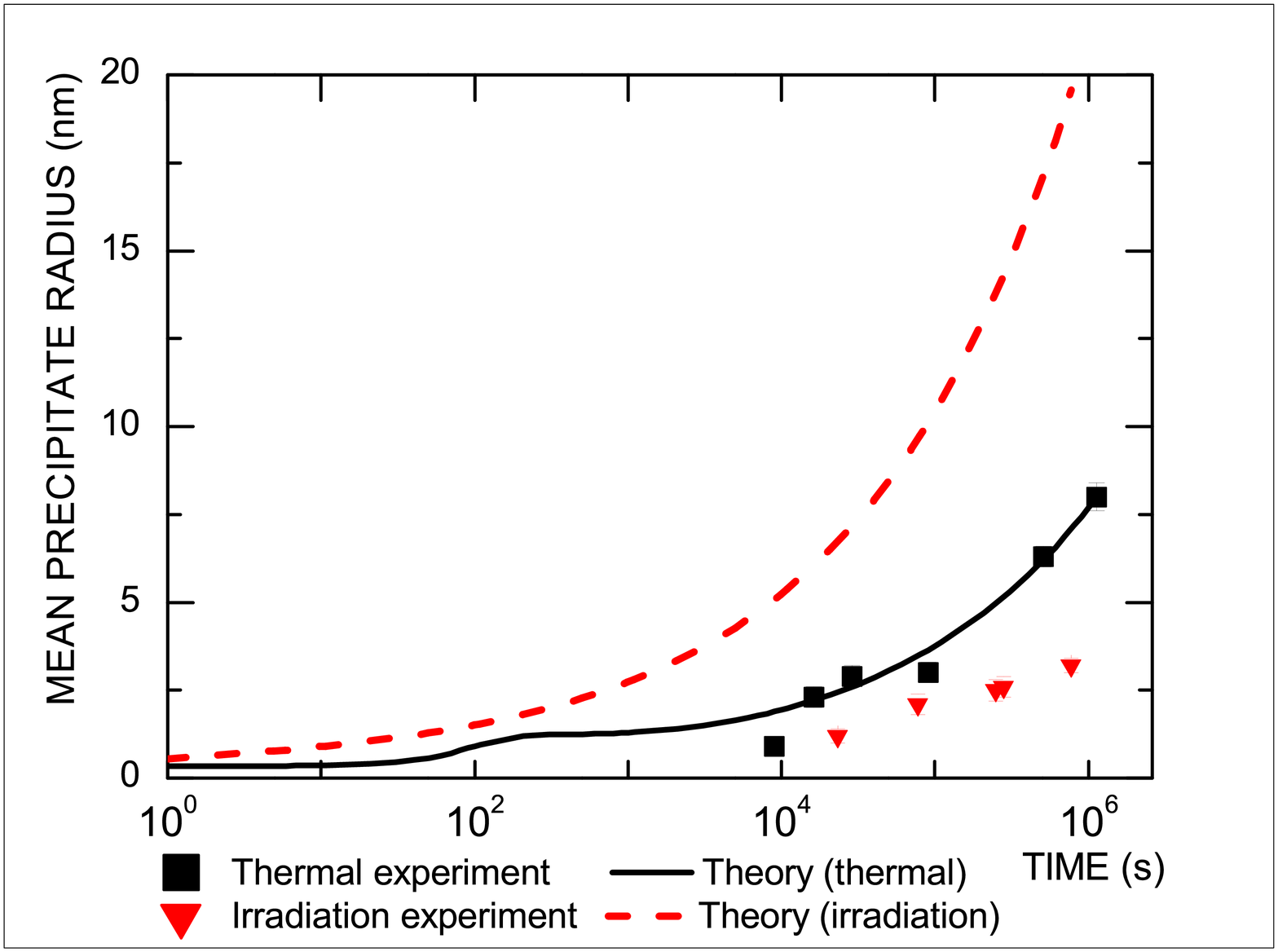}
\caption{Time evolution of the $Cu$ precipitate mean radius under irradiation at $563 K$ vs. experimental data \cite{Ref17}.}
\label{fig:5}  
\end{figure}
\begin{figure}
\center
\includegraphics[width=8cm]{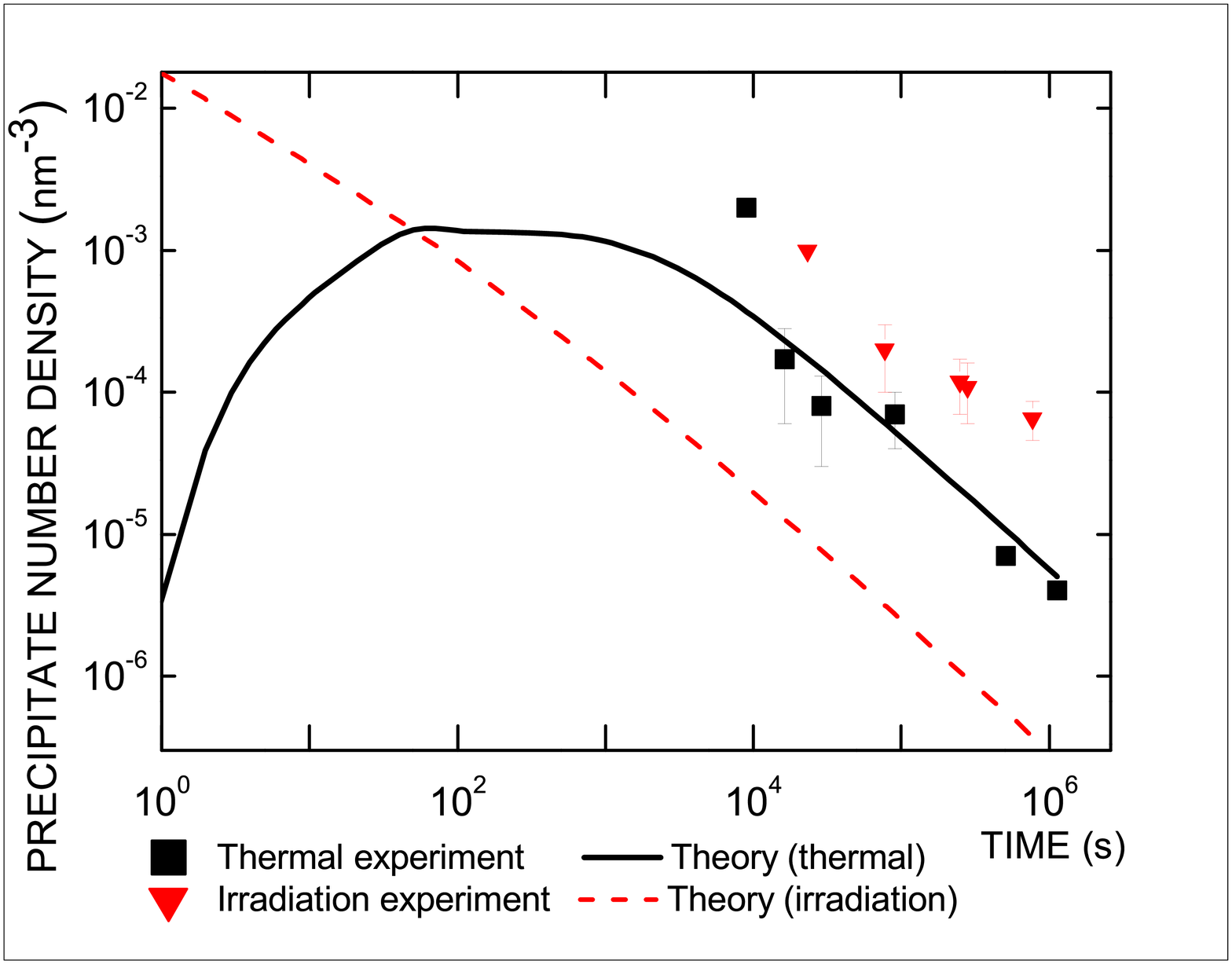}
\caption{Time evolution of the $Cu$ precipitate concentration under irradiation at $563 K$ vs. experimental data \cite{Ref17}.}
\label{fig:6}  
\end{figure}

Irradiation conditions in the ref.~\cite{Ref17} are as follows: temperature, $T_{irr} = 563 K$, displacement rate
$K = 2\times10^{-9} dpa/s$, total irradiation time $t_{irr} = 7.75\times10^5 s$, which corresponds to the total irradiation dose of $1.55\times10^{-3}$ displacements per atom (dpa).
\begin{figure}
\center
\includegraphics[width=8cm]{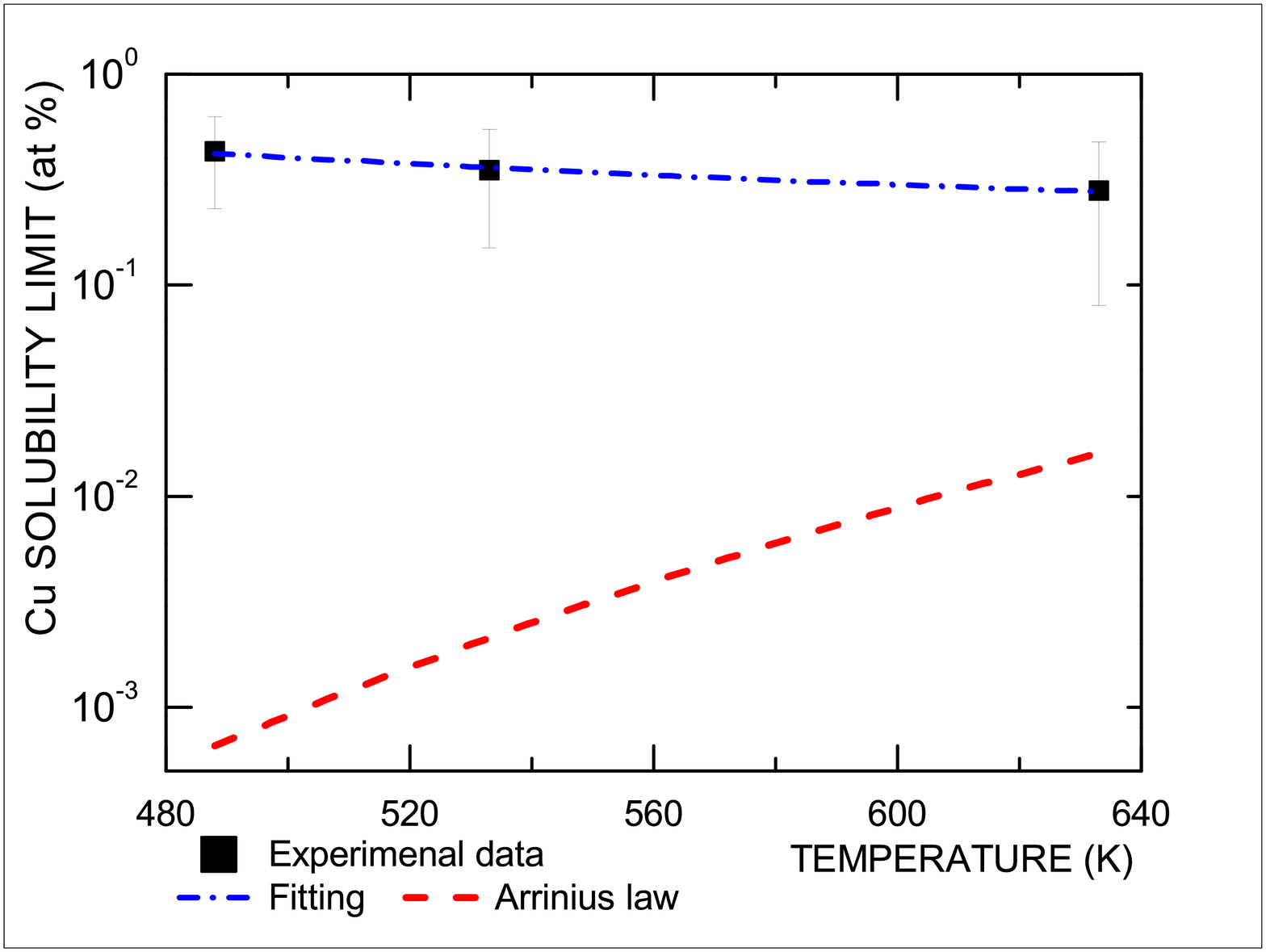}
\caption{Temperature dependence of the copper solubility limit under irradiation \cite{Ref17} vs. Arrhenius law.}
\label{fig:7}  
\end{figure}

It can be seen in Figs.~\ref{fig:5}~and~\ref{fig:6}, that the classical theory approach greatly overestimates the growth rate of the precipitate mean radius, and underestimates their concentration. As mentioned in the introduction, this discrepancy is a natural consequence of the main assumption of the classical theory on the solubility limit of copper being determined by Arrhenius formula (thermodynamic assumption). Since this assumption is used by all existing precipitation models, their results could not be validated by experimental data on the radiation-induced solubility limit, as it becomes evident from comparison between theoretical and experimental values presented in
Fig.~\ref{fig:7}. This discrepancy was actually admitted (as a puzzle) by the authors \cite{Ref17}, and it will be shown to be a natural consequence of the radiation-induced increase of the solubility limit due to the interaction of precipitates with irradiation-induced quodons. Below we use the copper diffusivity data from the first line in the Table~\ref{tab:Param} as the best fit to experimental data for thermal annealing.
\subsection{Quodon-induced solubility limit change}
\label{sec:3.2}
In this section, non-equilibrium thermodynamics of ``quodon gas'' produced by irradiation is taken into account in the evaluation of the solubility limit under irradiation.

Consider a free energy, ${\rm\Phi}^{th}_{Cu}(n, c)$ of a system consisting of an extended defect ($Cu$ precipitate) of n atoms and a diluted solid solution of point defects -- $Cu$ atoms in our case, with a concentration $c << 1$:
\begin{equation}\label{eq:23}
{\rm\Phi}^{th}_{Cu}(n, c) = \left(N_0 - n\right) \left(\phi^m_{Cu} + k_B T \ln c\right) +
4\pi\gamma b^2 n^{2/3} + n \phi^p_{Cu}
\end{equation}
where $N_0$ is the total number of point defects in the system, $\phi^m_{Cu}$ is the Gibbs potential one of that in the matrix, $\phi^p_{Cu}$ is the free energy of a defect in the precipitate, and $\gamma$ the interfacial energy. Minimization of the free energy with respect to $n$ variation, $\partial{\rm\Phi}^{th}_{Cu}/\partial n = 0$,
is known to result in the classical \underline{thermal equilibrium} concentration of point defects, which depends exponentially on $T$ and $n$:
\begin{equation}\label{eq:24}
\begin{array}{rcl}
&&C^{th}_{Cu}(T, n) = \exp\left(-\displaystyle\frac{{\rm\Delta}\phi^{mp}}{k_B T}\right)\exp\left(\beta n^{-1/3}\right), \\[2.0ex]
&&{\rm\Delta}\phi^{mp} = \phi^m_{Cu}-\phi^p_{Cu} > 0, \quad \quad
\beta = \displaystyle\frac{8\pi b^2\gamma}{3 k_B T},
\end{array}
\end{equation}
and defines a thermodynamic solubility limit of point defects, $C^{th}_{Cu0}(T)$:
\begin{equation}\label{eq:25}
C^{th}_{Cu}(T, n\rightarrow\infty) \longrightarrow C^{th}_{Cu0}(T).
\end{equation}
Here the difference ${\rm\Delta}\phi^{mp}$ is the \underline{minimum work} done by the system while transferring a poin defect from a solution to the precipitate under constant volume and temperature, which is usually referred to as the \underline{dissolution activation (free) energy}. It is defined as 
\begin{equation}\label{eq:26}
{\rm\Delta}\phi^{mp} = {\rm\Delta}E^{mp} - T {\rm\Delta}S^{mp},
\end{equation}
where ${\rm\Delta}E^{mp}$ is the  height of potential barrier (see Fig.~\ref{fig:6}), and ${\rm\Delta}S^{mp}$ is the entropy change due to point defect transfer. 

Under irradiation, free energy of the system is reduced due to fluctuations of the potential landscape near the interface region caused by the scattering of quodons at the interface (Fig.~\ref{fig:2}), which \underline{reduces} the minimum work done by the system while transferring a point defect from a solid solution to the precipitate, as has been demonstrated in section~\ref{sec:2.2}:
\begin{equation}\label{eq:27}
{\rm\Delta}\phi^{mp} = {\rm\Delta}\phi^{mp}_{th} - {\rm\Delta}\phi_q(F_{irr}, T), \quad
{\rm\Delta}\phi_q(F_{irr}, T) = \frac{V^2_q\omega_q(F_{irr}, T)\tau_0}{k_B T},
\end{equation}
where ${\rm\Delta}\phi^{mp}_{th}$ designates free energy of the system without irradiation and $F_{irr} > 0$ is a flux of impinging particles.

Accordingly, instead of thermodynamic expression~(\ref{eq:24}) one has to consider a quodon-modified 
\underline{dynamic equilibrium} concentration of point defects, which should depend on the irradiation flux that generates quodons:
\begin{equation}\label{eq:28}
C^q_{Cu}(F_{irr}, T, n) = C^{th}_{Cu}(T, n)\exp\left(\frac{{\rm\Delta}\phi_q(F_{irr}, T)}{k_B T}\right),
\end{equation}
which will give us the quodon-modified solubility limit , $C^q_{Cu0}(F_{irr}, T)$, as follows 
\begin{eqnarray}
&&C^q_{Cu0}(F_{irr}, T) = C^{th}_{Cu0}(T)\exp\frac{V^2_q\omega_q(F_{irr}, T)\tau_0}{(k_B T)^2}, \label{eq:29} \\ 
&&\omega_q(F_{irr}, T) = \frac{K_q\left(F_{irr}l_q(T)\right)W_a}{wb},   \label{eq:30}  
\end{eqnarray}
where $W_a = 4\pi R^2 b$ is the effective activation volume for the quodon-induced modification of the solubility barrier.

\begin{figure}
\center
\includegraphics[width=8cm]{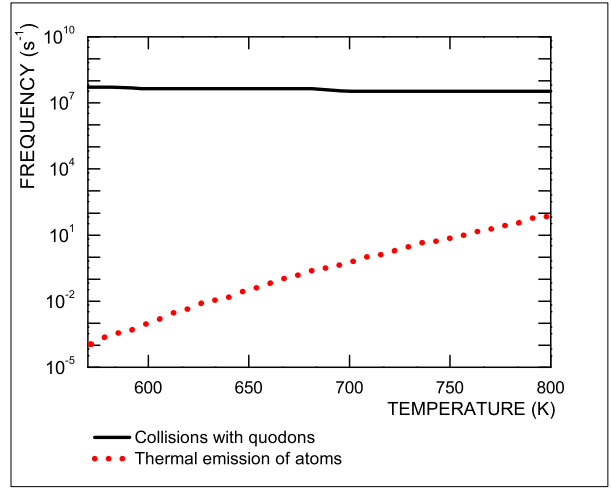}
\caption{Frequency of collisions between a precipitate and quodons vs. frequency of atom thermal emission.}
\label{fig:8}  
\end{figure}

As shown in Fig.~\ref{fig:8}, the frequency of collisions between a precipitate and quodons is about 
$10^7 {\rm s^{-1}}$, which greatly exceeds the frequency of thermal desorption of copper atoms into the matrix, which ranges from $6\times10^{-5}$ to $22$ $\rm s^{-1}$ in the temperature interval under investigation:
$T = (563\div773 \rm K)$. This means that the interaction between quodon ``gas'' and structural defects
(e.g. precipitates) can be viewed as that between an ideal molecular gas and Brownian particles. Statistical lows seem to be equally valid in both cases.
\begin{figure}
\center
\includegraphics[width=8cm]{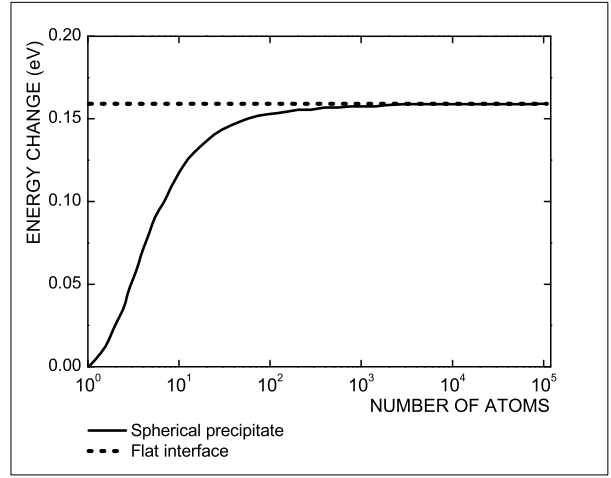}
\caption{Quodon-induced change of the dissolution free energy, ${\rm\Delta}\phi_q$, vs. the number of atoms in a precipitate.}
\label{fig:9}  
\end{figure}

Naturally, the quodon scattering at structural defects should depend on the defect size, ranging from low values for point defects (almost ``invisible'' for quodons) to the values estimated above for large extended defects.  This physical picture can be modeled simply by assuming that ${\rm\Delta}E_q(F_e, T)$ depends on the precipitate size using the following {\itshape size factor} $fit(n)$:
\begin{equation}\label{eq:31}
{\rm\Delta}\phi_q(F_e, T, n) = {\rm\Delta}\phi_q(F_e, T)\times fit(n), \quad
fit(n) = 10^{-3}\times\left(1+\frac{2\times10^3}{1 + \exp(5/n)}\right),
\end{equation}

For example, at irradiation temperature $T = 563 \rm K$, one gets in this way the barrier decrease
${\rm\Delta}\phi_q(F_e, T, n) = 0.119\rm eV$ for a small precipitate, and 
${\rm\Delta}\phi_q(F_e, T, n) = 0.157\rm eV$ for a large precipitate, which coincides with that for a flat interfacial surface, ${\rm\Delta}\phi_q(F_e, T, \infty)$, which determines the solubility limit.

Fig.~\ref{fig:9} shows dependence of the quodon-induced change of the dissolution energy on the number of atoms in a precipitate, given by eq.~(\ref{eq:31}) as compared to the value for a flat inerface.

\subsection{Quodon-induced interfacial energy change}
\label{sec:3.3}
We will use an equation for the interfacial energy $\gamma$ similar to that calculated by Mathon et al \cite{Ref17} based on the Cahn-Hilliard theory adapted for taking into account of non-configurational entropy. In thermal case, the equation is as follows:
\begin{equation}\label{eq:32}
\gamma_0(T) = 0.54\frac{k_B T}{b^2}\left(\ln\frac{1}{C^{th}_{Cu0}(T)} - 2\right) = 
0.54\frac{k_B T}{b^2}\left(\frac{E^f}{K_B T} - \frac{{\rm\Delta}S^{non}}{k_B} - 2\right),
\end{equation}
where $C^{th}_{Cu0}(T)$ is the \underline{thermal equilibrium} solubility at a flat interface. Under irradiation, the equation should be modified to take into account the quodon interaction with the surface. Accordingly, for a flat interface one has 
\begin{equation}\label{eq:33}
\begin{array}{rcl}
\gamma_0(F_{irr}, T) = &&
0.54\displaystyle\frac{k_B T}{b^2}\left(\ln\displaystyle\frac{1}{C^q_{Cu0}(F_{irr}, T)} - 2\right)
= \\[2.0ex]
&&0.54\displaystyle\frac{k_B T}{b^2}\left(\frac{E^f - {\rm\Delta}E_{q0}(F_{irr})}{K_B T} - \displaystyle\frac{{\rm\Delta}S^{non}}{k_B} - 2\right),
\end{array}
\end{equation}
where $C^q_{Cu0}(F_{irr}, T)$ is the quodon-induced solubility limit.

\begin{figure}
\center
\includegraphics[width=11.5cm]{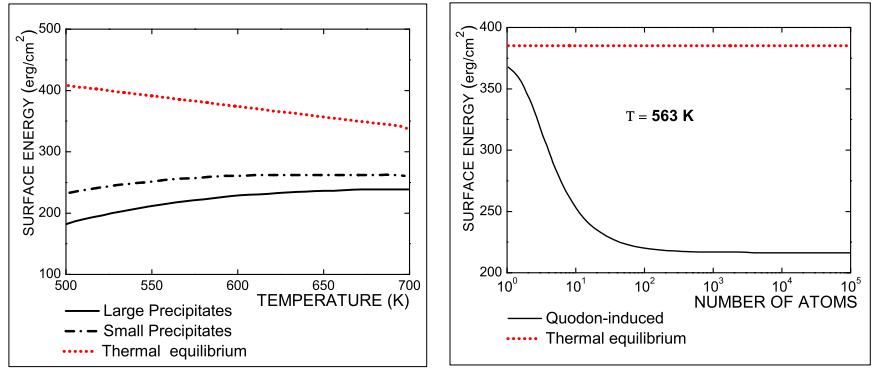}
\caption{Quodon-induced interface energy as a function of temperature and the number of atoms in the precipitate.}
\label{fig:10}  
\end{figure}

\begin{figure}
\center
\includegraphics[width=11.5cm]{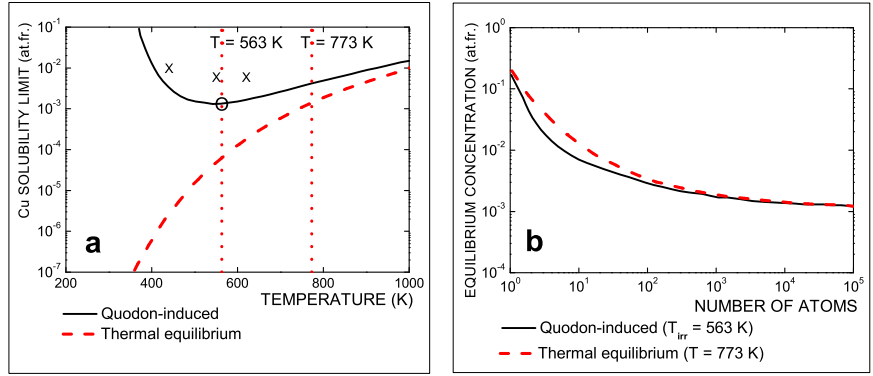}
\caption{\textbf{(a)} Thermal equilibrium and quodon-induced solubility limits of $Cu$ in a $Fe$ matrix; \textbf{(b)} equilibrium concentrations at precipitates of different sizes; \textbf{x} and \textbf{O} are the experimental data \cite{Ref17} for electron irradiation of $FeCu$ alloy.}
\label{fig:11}  
\end{figure}

In order to take into account size dependence of the quodon-precipitate scattering, we will modify eq.~(\ref{eq:33}) similar to eq.~(\ref{eq:31}), and introduce a size-dependent interfacial energy:
\begin{equation}\label{eq:34}
\gamma_q(n) = 0.54\frac{k_B T}{b^2}\left(\frac{E^f - fit_\gamma(n) {\rm\Delta}E_{q0}(F_{irr})}{K_B T} - \frac{{\rm\Delta}S^{non}}{k_B} - 2\right),
\end{equation}
where the size factor $fit_\gamma(n)$ is different from that in eq.~(\ref{eq:31}):
\begin{equation}\label{eq:35}
fit_\gamma(n) = 10^{-1}\times\left(1+\frac{2\times10^1}{1 + \exp(5/n)}\right).
\end{equation}
The resulting dependence of the interface energy on temperature and precipitate size is shown in Fig.~\ref{fig:10}. In this way one can evaluate the concentration of $Cu$ in \underline{dynamic equilibrium} with precipitates of different sizes at different temperatures and irradiation conditions, as demonstrated in Fig.~\ref{fig:11}.


One can see that our new concept is able to describe both high copper solubility limit (compared to the thermal value) in agreement with that measured in~\cite{Ref17} (Fig~\ref{fig:11}a) and quite comparable ``local equilibrium'' concentrations near small copper clusters at thermal annealing at $773\rm K$ and under irradiation at $563\rm K$ (Fig.~\ref{fig:11}b), which actually determine the precipitate critical size and nucleation rate. So the diffusivities at these two experimental setups should be also comparable in order to get reasonable agreement with observed nucleation and growth rates.

\subsection{Diffusivity of Cu under irradiation}
\label{sec:3.4}

Diffusivity of $Cu$ is determined by the vacancy mechanism and it can be written as follows:
\begin{equation}\label{eq:36}
D^{irr}_{Cu} = 4 b^2 \upsilon_{Cu} C^{irr}_V
\exp\left(\frac{{\rm\Delta}S^m_{Cu,V}}{k_B}\right)
\exp\left(-\frac{{\rm\Delta}E^m_{Cu,V}  - E^b_{Cu,V}}{k_B T}\right),
\end{equation}
where $\upsilon_{Cu}$ is the frequency factor, ${\rm\Delta}S^m_{Cu,V}$ is the entropy factor,
${\rm\Delta}E^m_{Cu,V}$ and $E^b_{Cu,V}$ are the $Cu$-vacancy migration and binding energies, respectively.

\begin{figure}
\center
\includegraphics[width=11.5cm]{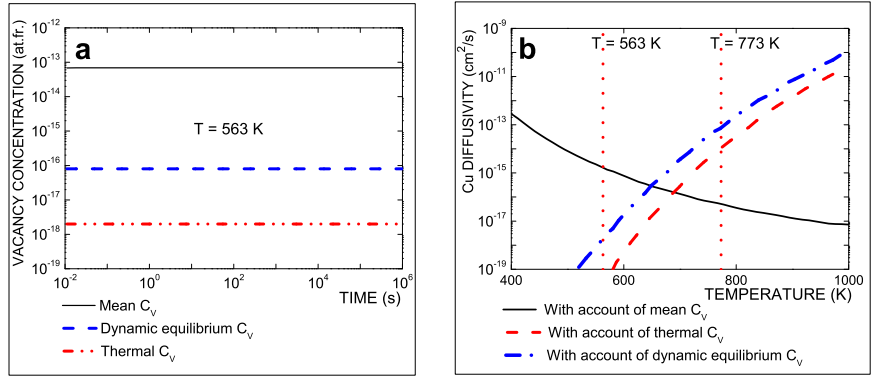}
\caption{Equilibrium (dynamic and thermal) and radiation-induced concentration of vacancies \textbf{(a)} and and diffusivities of $Cu$ in a $Fe$ matrix \textbf{(b)} calculated for electron irradiation  with account of different vacancy mechanisms \cite{Ref17}.}
\label{fig:12}  
\end{figure}

It appears that the result depends crucially on the mean vacancy concentration, $C^{irr}_V$, which is given by the balance of their production and annihilation in the bulk and at extended defects \cite{Ref1, Ref28}:
\begin{equation}\label{eq:37}
\begin{array}{rcl}
C^{irr}_V(F_e, E_e, T) &=&
\displaystyle\frac{K_d(F_e, E_e)}{k^2_p + k^2_d} + C^q_V(F_{irr}, T),
\\[2.0ex]
k^2_{ED} &\approx& 4\pi N_p R_p + \rho_d,
\end{array}
\end{equation}
where $K_d(F_e, E_e)$ is the displacement rate, $k^2_{ED}$ are the dislocation and precipitate sink strength for vacancies determined by the precipitate concentration $N_p$ and mean radius $R_p$, and by the dislocation density, $\rho_d$; $C^q_V$ is the mean dynamic equilibrium concentration of vacancies at all extended defects in the system, which (similar to the equilibrium concentration of $Cu$) depends both on temperature and irradiation flux \cite{Ref28}, as shown in Fig.~\ref{fig:12}a.

Without irradiation, $C^q_V(0, T)$ corresponds to the thermal equilibrium concentration shown in Fig.~\ref{fig:12}a for comparison. Under irradiation it increases due to the radiation-induced vacancy emission from extended defects, but this effect is overshadowed by the vacancy production in the bulk and annihilation at extended defects. The main contribution to vacancy annihilation comes from precipitates, since their sink strength (measured experimentally in~\cite{Ref17}) is higher than that of the dislocations by several orders of magnitude under these experimental conditions. Precipitates behave under irradiation similar to gas bubbles \cite{Ref28} adjusting the number of vacancies in them adiabatically to the number of copper atoms due to the positive feedback from the misfit stress on the bias of precipitates for absorption of interstitial atoms. As a result, precipitates act as strong recombination centres for vacancies and interstitial atoms.

As shown in Fig.~\ref{fig:12}b, the $Cu$ diffusivity under irradiation at $563 \rm K$ is determined by the mean vacancy concentration and become comparable with that under thermal annealing at $773 \rm K$. 
The ratio of $D^{th}_{Cu}(773)/D^{irr}_{Cu}(563) \approx 6.9$, seems to be in excellent agreement with a factor of $\sim 5$ more rapid thermal evolution of precipitates at $773 \rm K$ as compared to that under irradiation at $563 \rm K$ observed by Mathon et al \cite{Ref17}.

In the following section we will test our predictions by modeling the evolution of $Cu$ precipitates and the matrix concentration of $Cu$ under electron irradiation \underline{with and without} account of the radiation-induced quodons and compare the results with experimental data.

\subsection{Evolution of Cu precipitates and the matrix concentration of Cu under electron irradiation}
\label{sec:3.5}
\begin{figure}
\center
\includegraphics[width=11.5cm]{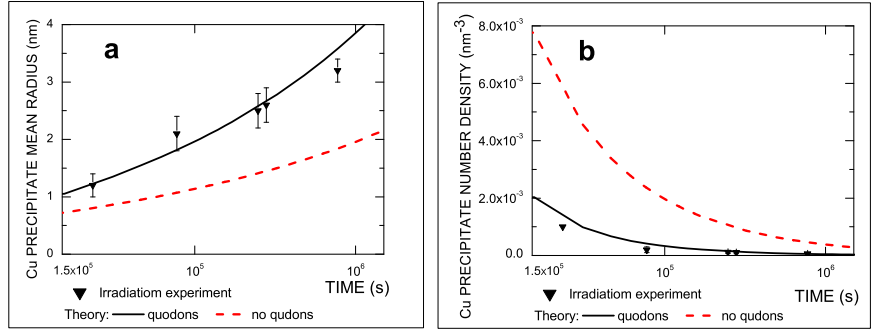}
\caption{Time evolution of the $Cu$ precipitate: \textbf{(a)} -- mean radius  and \textbf{(b)} -- concentration  vs. experimental data \cite{Ref17}.}
\label{fig:13}  
\end{figure}

\begin{figure}
\center
\includegraphics[width=8cm]{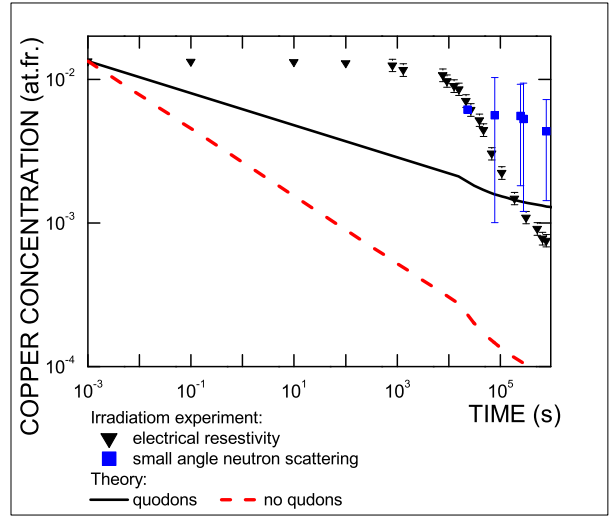}
\caption{Time evolution of the concentration of copper atoms under thermal annealing and irradiation. Experimental data by Mathon et al \cite{Ref17} obtained by electrical resistivity method are shown by triangles together with results obtained  by small angle neutron scattering shown by squares.}
\label{fig:14}  
\end{figure}

The evolution of $Cu$ precipitates and the matrix concentration of $Cu$ under electron irradiation have been modeled using parameters from Table~\ref{tab:Param}. The results are compared with results of the classical model (without quodon-induced effects) and with experimental data \cite{Ref17} in Figs. \ref{fig:13} and \ref{fig:14}.

One can see that the classical model disagrees strongly with experimental data on all precipitation parameters, and this discrepancy is especially pronounced for the matrix concentration of copper (Fig.~\ref{fig:14}). In contrast, the quodon-modified model describes quite well both the evolution of precipitates and the matrix concentration of copper measured by different methods. The latter fact seems to be of particular importance, since it reflects the principal difference between the two concepts, namely, the ``classical'' and the new one, in relation to the mechanisms of production of Schottky defects.
In the classical theory they are assumed to be emitted by extended defects exclusively due to thermal fluctuations, which, in our language, are driven by phonons. In the modified rate theory we have taken into account essentially \underline{athermal fluctuation} mechanisms based on the interaction of extended defects with radiation-induced quodons. It should be noted that these mechanisms have been considered in the previous works
\cite{Ref1}, \cite{Ref3}, \cite{Ref4}, \cite{Ref5}, \cite{Ref6}, \cite{Ref28} dealing with radiation-induced emission of another kind of Schottky defects, namely, vacancies.  Similar to the solute atoms, the dynamic equilibrium concentration of vacancies has been evaluated (in the framework of ``ballistic'' concept) to be much higher than thermal equilibrium values. However, it is rather difficult technically to measure the dynamic equilibrium concentration of vacancies directly by electric resistivity measurements due to the lack of sufficient accuracy of this method. In contrast, the equilibrium concentration of solute atoms, such as $Cu$, can be measured directly at the final stage of the precipitate evolution, and this has been done by Mathon et al \cite{Ref17}, which makes their work particularly relevant for experimental discrimination between the classical and quodon-modified models.

\section{Summary}
\label{sec:4}
From a methodological side, the proposed concept of the radiation-induced ``gas'' of quodons offers a new insight on the radiation-induced processes in solids. It appears that 
{\itshape quodons are the transient form of the heat generation under irradiation} that subsequently transfers energy to phonons. The quodon gas may be a powerful driver of the chemical reaction rates under irradiation, the strength of which exponentially increases with irradiation flux and may be comparable with strength of the phonon gas that exponentially increases with temperature. Phonons obey the lows of equilibrium thermodynamics, such as the minimization of a system free energy, and the latter could not be defined for a crystal under irradiation in the classical framework. The proposed method of the free energy modification by taking into account non-equilibrium fluctuations of the potential landscapes for chemical reactions offers a self-consistent way for description of the radiation-induced reactions and evolution of the microstructure in the nuclear materials.

A detailed description of the rate theory modelling of the precipitation kinetics has been presented and applied to describe the precipitation of copper in iron matrix under thermal annealing and under irradiation. The modelling has been performed both in the framework of a {\itshape classical} rate theory and the {\itshape modified} rate theory, which takes into account non-equilibrium fluctuations driven by the ``gas'' of radiation-induced quodons. The classical theory was shown to disagree strongly with experimental data on all precipitation parameters. In contrast, the quodon-modified theory describes quite well both the evolution of precipitates and the matrix concentration of copper measured by different methods in $FeCu$ binary alloys under electron irradiation.

\section*{Acknowledgments}

This study has been supported by the STCU grants \#5228;  \#5497, PNNL EED LDRD, and the U.S. Department of Energy, Office of Fusion Energy Sciences under contract DE-AC06-76RLO 1830.

Helpful discussions with Professors Vladimir V. Hizhnyakov and Juan F. R. Archilla are gratefully acknowledged. 



\printindex
\end{document}